\pdfoutput=1 
\documentclass{llncs}

\usepackage{makeidx} 
\usepackage{graphicx}
\usepackage{times}
\usepackage{amstext,amssymb,amsmath}
\usepackage{color}

\newcommand{\cancel}[1]{}

\begin{document}
\frontmatter          
\pagestyle{headings}  
\mainmatter              
\title{Complementary Space for Enhanced Uncertainty and Dynamics Visualization}
\titlerunning{Complementary Space for Visualization}  
%
\author{Chandrajit Bajaj \and Andrew Gillette \and Samrat Goswami \and Bong June Kwon \and Jose Rivera}

\authorrunning{Bajaj et al.}   
%
\tocauthor{Chandrajit Bajaj, Andrew Gillette, Samrat Goswami, Bong June Kwon, Jose Rivera}

\institute{Center for Computational Visualization, University of Texas at Austin, Austin, TX 78712, USA,\\
\texttt{http://cvcweb.ices.utexas.edu/ccv/}
}

\maketitle              


\begin{figure}
\centering
\includegraphics[width=3in]{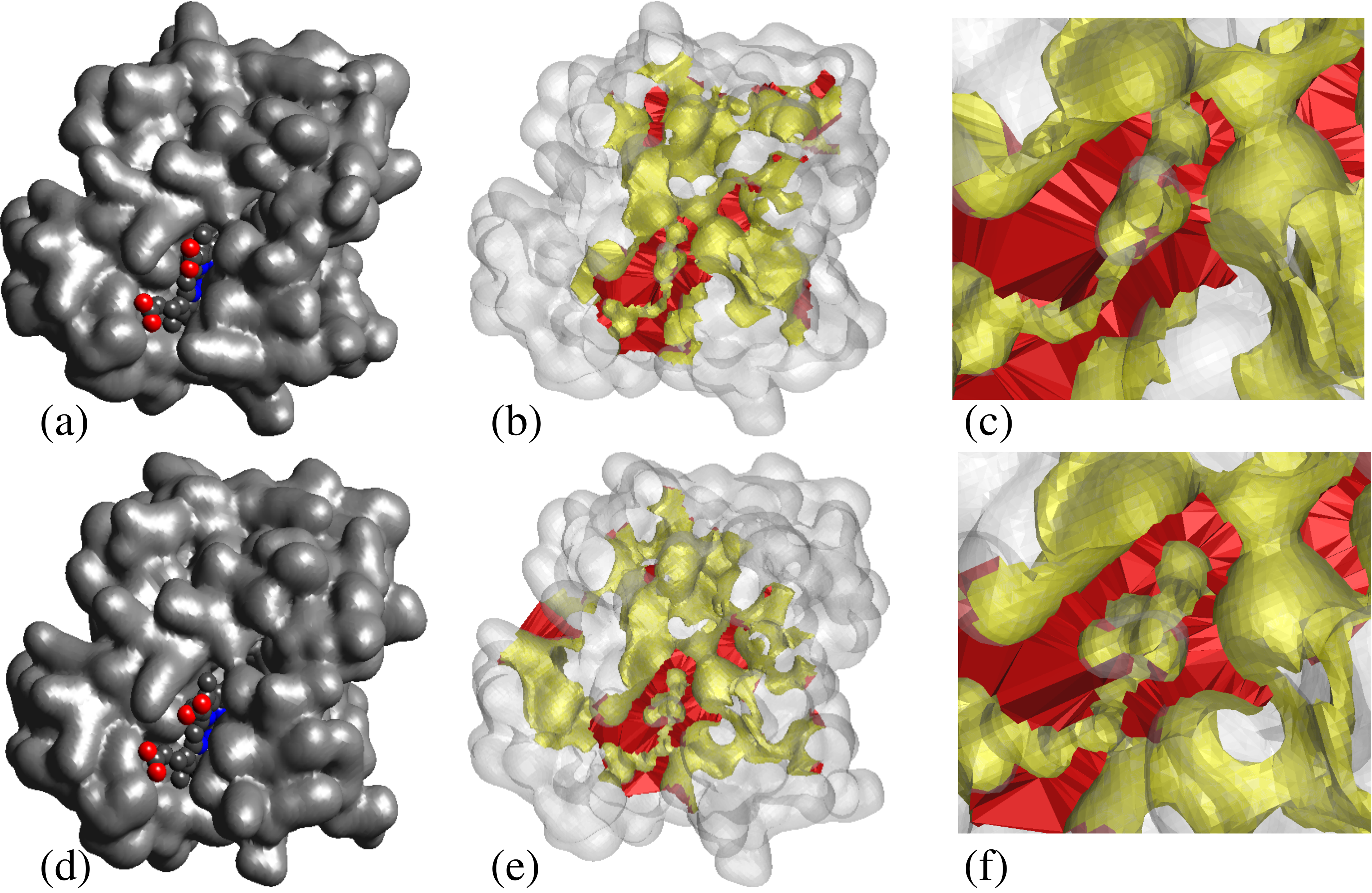}
\caption{
Visualizations of the hemoglobin molecule undergoing dynamic deformation as oxygen binds to it.  \textbf{(a)} A primal space visualization of the first time step with the heme group identified.  \textbf{(b)} Visualization of the complementary space of the first time step shows the geometry of the interior.  The surface has been made transparent, revealing a large tunnel through the surface (yellow) with many mouths (red).  \textbf{(c)} Zooming in on the heme group reveals the structure of space around it while oxygen is bound. \textbf{(d-f)} The corresponding images of (a-c) for the final time step.  Complementary space has changed dramatically both in the interior volume and near the heme group, though this is not evident from the primal space visualizations.  Comparing (c) and (f), we observe that the connectivity of the mouth of the tunnel near the heme group has changed, illustrating the time-dependency of the topological features of complementary space.  We quantify and discuss this example further in Section \ref{sec:hemo}.}
\label{fig:hemVideo_viz}
\end{figure}

\section{Introduction}
\index{complementary space}
\index{Molecular dynamic}
\index{Molecular simulation}
\index{Topological data analysis}
\index{Tracking accuracy}
\index{Uncertainty}

Many computational modeling pipelines for geometry processing and visualization focus on topologically and geometrically accurate shape reconstruction of ``primal'' space, meaning the surface of interest and the volume it contains.  Certain features of a surface such as pockets, tunnels, and voids (small, closed components) often represent important properties of the model and yet are difficult to detect or visualize in a model of primal space alone.  It is natural, then, to consider what information can be gained from a model and visualization of complementary space, i.e. the space exterior to but still ``near'' the surface in question.  In this paper, we show how complementary space can be used as a tool for both uncertainty and dynamics visualizations and analysis.

Uncertainty visualizations aim to elucidate the accuracy of a model by visibly identifying and subsequently quantifying potential errors in the model.  For surface and volume models, drawing attention to topological errors is of particular importance as they represent a more fundamental inaccuracy in shape than geometrical errors.   Topological errors include the presence of an unwanted tunnel in a surface, the absence of a desired tunnel, and the existence of small extraneous components.  We also consider errors related to the existence or absence of ``pockets'' in a surface as topological errors; pockets on a primal space surface correspond to components in complementary space and the number of components of a space is a basic topological property.  As we will discuss, it is important to determine if topologically distinct models occur within the inherent uncertainty bounds of a model and complementary space provides a natural means for visualizing uncertainty in topological structure.

Dynamics visualizations bring shape models to life by showing conformational changes the model may undergo in the application context.  In this case, there is uncertainty not only in the particularities of the shape at any time step of the dynamics, but also in the plausibility of the simulated movement as a whole.  The creation of a tunnel, the collapse of a pocket into a void, or the changing geometry of the interior of a tunnel may be highly relevant to understanding a dynamic situation and assessing its likelihood of simulating reality.  By visualizing and as necessary quantifying complementary space at each stage, we better understand how our models need to be refined.  In Figure \ref{fig:hemVideo_viz} and the included videos, we show an example of a dynamical situation where complementary space visualization and quantification aids in understanding.

The specific problems we address in this paper touch on a variety of situations where modeling is sensitive to topological errors in primal or complementary space.  In each case, we discuss how complementary space modeling provides a natural method for processing, visualizing, and managing such errors.  Our methods are inspired by challenges encountered in our work with biological models and hence most of our illustrated examples use actual biological data from public sources and our academic collaborators.  Nevertheless, the computational techniques we describe here are useful in a variety of settings including creation of exploded assembly images and videos, quantification of complementary space features in CAD models, and visualization of potential errors in any image-based modeling scheme.

We conclude this section with a brief review of related work in complementary space modeling.  In Section \ref{sec:bkgd} we explain the relevant theory behind our two approaches to complementary space modeling and fix notation.  In Section \ref{sec:approach} we describe particular approaches to complementary space solutions we have implemented in our lab and show some results.  In Section \ref{sec:conc} we conclude and discuss future work.

\paragraph{Related Work}

We discuss three approaches to complementary space modeling: alpha shapes, surface propagation, and Morse theory for distance functions.

The notion of alpha shapes was developed by Edelsbrunner et al. \cite{EFL1998} in the context of molecular modeling with the aim of pocket detection.  The method uses the Delaunay diagram on the atomic centers of a molecule and decomposes space into Delaunay tetrahedra near the molecule and into unbounded regions away from it.  By assigning a flow across Delaunay faces based on the radius of nearby atoms, the authors distinguish between those finite tetrahedra belonging to the alpha shape (i.e. the molecule) and those belonging to a pocket.  This method has been implemented and applied to a number of proteins with some success \cite{LEW1998}.  In instances of broad pocket mouths, however, it is difficult for this method to detect the pocket and its coarse treatment of the molecular model leaves geometrical refinement to be desired.  It is also unclear how this approach could be applied to non-molecular models.

The method of surface propagation represents a surface implicitly as the zero level set of a distance function.  The surface is subjected to an evolution equation first introduced by Osher and Sethian \cite{OS1988}.  To turn this from an analytical theory to a computable implementation, Sethian developed a fast level set marching method \cite{S1996} for propagating the given surface in the outward normal direction.  Zhang and Bajaj \cite{ZB2008} have used this method to propagate a molecular surface outward until the surface has the topology of a sphere and then back inward for an equal length of time.  Pockets are then defined as those points exterior to the initial surface and interior to the final surface.  This method has been implemented in our publicly available software package TexMol \cite{texmol}.
\index{critical point}
\index{Morse complex}
\index{Morse-Smale complex}
\index{Morse theory}

A more general approach to complementary space modeling uses the Morse complex for distance functions.  The Morse complex canonically decomposes a space relative to its features, as identified by the critical points of a chosen function.  Cazals, Chazal and Lewiner have computed the Morse complex for the Connolly function on 2-manifolds \cite{CCL2003} which has shown some promise as a primal space method for shape analysis.  Natarajan and Pascucci have used the Morse complex for aid in visualization of cryo-EM data \cite{NP2005}.  The Morse complex for a distance function can be approximated by inducing a flow on the Voronoi diagram of a point sample of the surface, as was characterized by Edelsbrunner \cite{E2002} and Giesen and John \cite{GJ2003}.  The distance function has been used in a variety of applications, including image feature identification \cite{C2003,YB2004}, stable medial axis construction \cite{CL04}, object segmentation and matching \cite{DGG2003}, annotation of flat and tubular features \cite{GDB2006}, and detection of secondary structural motifs in proteins \cite{BG2006}.  In previous work, we have shown how the distance function can also be used to detect tunnels and pockets \cite{BGG2007} for the purpose of surface curation.  In this paper, we show how these techniques can be used for the much broader purpose of uncertainty and dynamics visualizations.

\section{Background and Notation}
\label{sec:bkgd}

We have implemented two complementary space modeling techniques: an out-and-back surface propagation method and a Morse complex based method.  We describe each in detail below and compare them at the end of this section.
\index{Surface mesh}

For out-and-back surface propagation, we begin with a meshed geometry of the initial surface $\Sigma$ and set $M$ to be a large, contractible, compact subset of ${\mathbb R}^3$ containing $\Sigma$ (e.g. a filled bounding box).  We define the distance function $h_\Sigma$ by
\begin{equation}\label{eqn:hsig}
h_{\Sigma}:M\rightarrow{\mathbb R},\quad x\mapsto\pm\inf_{p\in\Sigma}||x-p||
\end{equation}
where the sign of the output is determined by the location of the input $x$ relative to $\Sigma$ (inside or outside). We represent $\Sigma$ implicitly as the zero level set $\{x\in{\mathbb R}^3:h_\Sigma(x)=0\}$.  The evolution equation for the surface is given by
\[\left\{\begin{array}{rcl} \phi_t + F|\nabla\phi| & = & 0,\\ \phi(x,0) & = & h_\Sigma(x).\end{array} \right.\]
Here, $F$ is a speed function in the normal direction, usually chosen to depend on the curvature.  To define pockets, we use this method to propagate a molecular surface outward  with constant speed ($F\equiv 1$) until some time $t$ when the surface has the topology of a sphere.  This surface is then propagated with speed $F\equiv -1$ also for time $t$, creating a final surface $\Sigma'$.  Pockets are then defined as those points exterior to $\Sigma$ and interior to $\Sigma'$. This method has been implemented in our publicly available software package TexMol \cite{texmol}.

The second approach we use for complementary space modeling employs the tools of Morse theory on the distance function $h_\Sigma$.  We define \index{critical point}\index{Triangulation}
\[h_P:M\rightarrow{\mathbb R},\quad x\mapsto\pm\min_{p\in P}||x-p||,\]
where  $\Sigma$ is the Delaunay diagram of a point sample $P$.  The function $h_P$ is easy to compute as opposed to $h_\Sigma$ which has no closed form in the general case.  The critical points of $h_P$ correspond to the intersection of Voronoi objects of Vor $P$ with their dual Delaunay objects in Del $P$.

This explicit description of the critical point structure of $h_P$ allows for the creation of a quickly computed discrete approximation of the Morse complex.  The complex is used to construct a geometry of complementary space by an algorithm we call \textsc{CompSpace}.  We described the details of this approach in detail in a previous issue of this series \cite{BGG2007} and do not repeat them here.  Once we have the geometry output by this method, we classify components with one mouth as pockets and components with two or more mouths as tunnels.  We can then tag the mouth, pocket interior, and tunnel interior simplicies different colors for visualization or compute their areas and volumes for quantification.

\paragraph{Comparison of the Methods}

We have implemented both methods described above and find that they are each useful in different circumstances.  The surface propagation method is ideal for detecting wide, shallow depressions on surfaces as the method detects regions of inward curvature.  In molecular modeling, such depressions occur at the active site for the binding of of a protein; in neuronal cell modeling, such depressions occur at synapses, the intracellular region between two adjacent cells across which a voltage signal is passed.  The Morse complex method excels in detecting deeper pockets and distinguishing tunnels from pockets as the Morse complex is laden with connectivity information.  We next look at a number of such examples using this approach.

\section{Applications}
\label{sec:approach}

Complementary space visualization can be used for model checking, error analysis, detection of topologically uncertain regions, topological preservation in model reduction, and dynamic deformation visualization, as we outline in the following subsections.

\subsection{Ion channel models}
\label{subsec:2bg9}
\begin{figure}
\centering
\includegraphics[width=\textwidth]{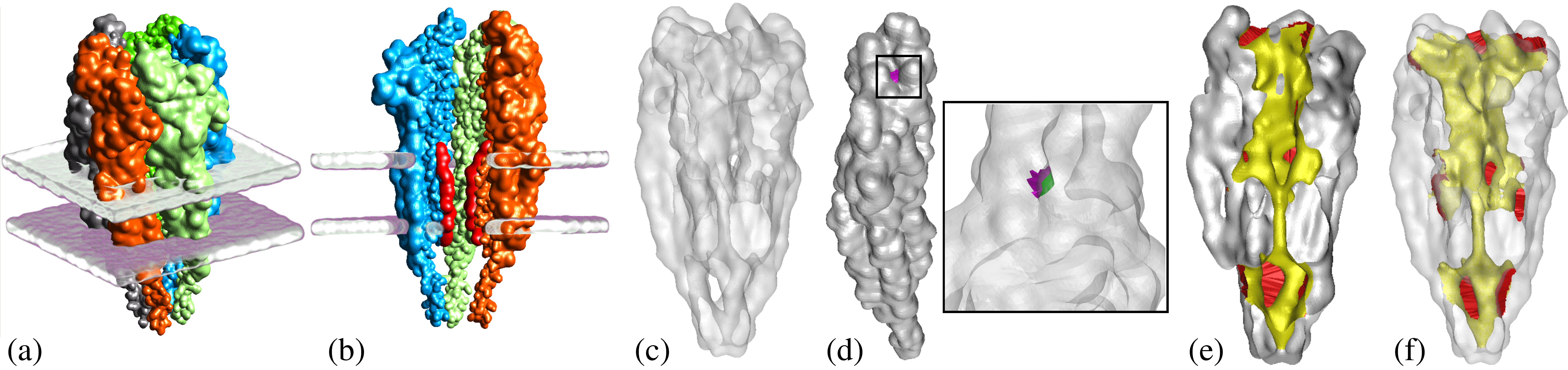}
\caption{Visualizations of the acetylcholine receptor molecule.  \textbf{(a)} The molecule is shown as it would sit embedded in a bilipid cell membrane (grey) with the five identical subunit colored for identification. \textbf{(b)}  A cut-away view of the same model showing where ions may pass through the center.  \textbf{(c)} A transparent view of the molecular surface.  \textbf{(d)} Each subunit contains a pocket where acetylcholine binds.  A complementary space view of the pocket interior (green) and its mouth (purple) are shown in a zoomed in view after the surface has been made transparent. \textbf{(e)}  A cut-away view of the surface with the interior of the tunnel (yellow) and its mouths (red) identified.  \textbf{(f)}  The same view as (c) with the tunnel geometry opaque, showing how it lies inside the surface.  Visualizing the complementary space tunnel reveals that the dimensions of the pore opening on the extracellular side are much larger than on the intracellular side.  This is less evident from the primal space visualizations.}
\label{fig:2BG9teaser}
\end{figure}

Ion channels are a cell's mechanism for regulating the flow of ions into and out of the cell.  They usually have two main structural confirmations: the `open' configuration, in which the tunnel through its center is wide enough to allow passage of the ions, and a `closed' configuration in which it is not.  We look at the acetylcholine receptor (PDB ID 2BG9) as a particular example of an ion channel.   This molecule is embedded in a cell membrane, as shown in Figure \ref{fig:2BG9teaser}a, and is a control mechanism for the flow of sodium and potassium ions into the cell.  It is made up of five homologous (in the biological sense) subunits.  A conformational change from closed to open occurs when acetylcholine, a small neurotransmitter ligand, docks into the five small pockets on the exterior of the molecule near the tunnel opening in the extracellular region.  When acetylcholine fills one of these pockets, it causes the attached chain subunit of 2BG9 to twist slightly.  The combined effect from rotations in all five chains is a widening of the mouth of the tunnel, akin to the opening of a shutter on a camera.

From this description of the action of the acetylcholine receptor, the importance of accurate complementary space topology becomes evident.  First, an accurate model of the channel must feature a tunnel passing completely through the length of the surface.  Put differently, the complementary space should include a connected component running the length of the molecule with mouths at opposite ends.  Such a requirement can be quickly verified by a complementary space visualization as shown in Figure \ref{fig:2BG9teaser}.  Furthermore, the diameter of this tunnel at its narrowest point should be within the range of biological feasibility, i.e. it should be wide enough to accommodate sodium and potassium ions in the open confirmation and narrow enough to block them in the closed confirmation.

To quantify properties of the tunnel, we use the geometries output by \textsc{CompSpace}.  The output of \textsc{CompSpace} is a mesh of the tunnel's interior surface, closed off by its mouths.  We compute this mesh for two models of the molecule - one in its `open' state and one in its `closed' state.  The enclosed volume in the open state is 86,657 $\text{\AA}^3$ and 60,045 $\text{\AA}^3$ in the closed state.  The mouth area in the open state is 5716 $\text{\AA}^2$ and 3290 $\text{\AA}^2$ in the closed state.  The minimum diameter in the open state is 5.9 $\text{\AA }$ and 8.0 $\text{\AA }$ in the closed state.  As expected, all the measurements - minimum tunnel diameter, mouth area, and enclosed volume - are all smaller in the closed state than in the open state.  We note that the change in the minimum diameter from the closed to open state is reasonable for accommodating ions whose width is a few angstroms.  We will augment this model in the future by incorporating electrostatic calculations of ion attraction and repulsion forces to further explain the gate-like ability of the channel.

An additional consideration for this model is its geometry at the ligand binding site.  In terms of complementary space, we expect a small pocket on each of the subunits such that its volume and mouth diameter are of plausible size compared to the acetylcholine molecule.   We show a visualization of the pocket in one subunit in Figure \ref{fig:2BG9teaser} d.  While such features are difficult to visualize and measure with a model based primal space, they much easier to detect and manipulate with a model based on complementary space.

\subsection{Ribosome models}\label{subsec:ribo}

\begin{figure}
\centering
\includegraphics[width=\textwidth]{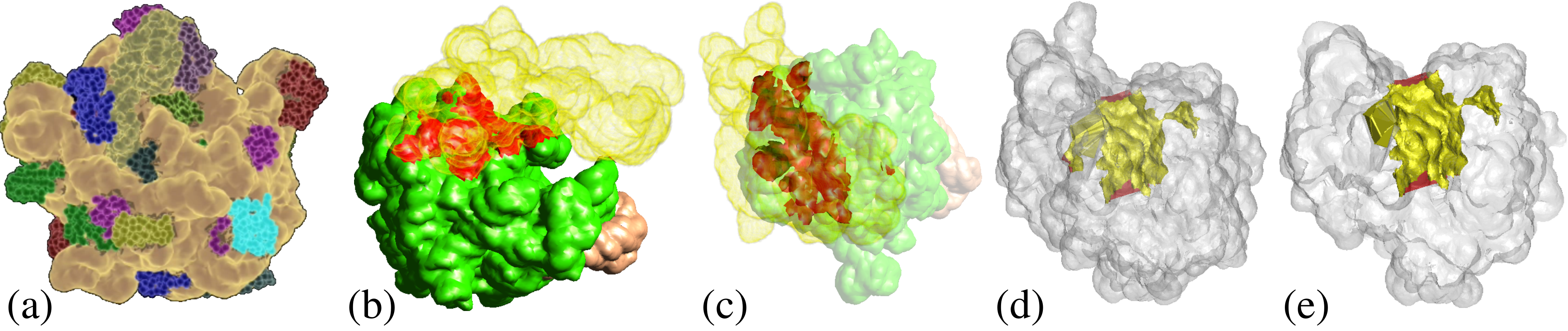}
\caption{Visualizations of the ribosome molecule. \textbf{(a)} The ribosome itself is made up of three RNA chains (brown) and dozens of proteins (various colors).  We define the contact area of each protein to be any portion of its surface lying within 4 $\text{\AA}$ of an RNA chain.  We compute these areas and use them to predict the order in which the proteins bind to the RNA chains. \textbf{(b)} Two of the RNA chains (green and brown) belong to the 50s subunit while the third (yellow) belongs to the 30s subunit.  We compute the contact area between these chains (red) to show how the subunits come together to form the protein assembly tunnel.  \textbf{(c)} A different view of the RNA chains gives a better view of the contact region but makes it difficult to see where the tunnel lies.  \textbf{(d)} We run \textsc{CompSpace} on a surface model that includes all the attached proteins and visualize the surface (transparent) along with the tunnel mouths (red) and interior (yellow).  \textbf{(e)} A cut-away view helps elucidate the intricate geometry surrounding the protein assembly region.  These types of complementary space visualizations and subsequent quantifications provide insight to open questions such as whether amino acid chains have room to begin folding before they exit the ribosome.}
\label{fig:ribosomeComp}
\end{figure}

The ribosome molecule provides another example of natural structural questions best answered with a complementary space model.  Ribosomes live inside cells and are the construction equipment for proteins made within the cell.  Proteins are assembled in a large tunnel that passes through the ribosome.  A copy of DNA data called mRNA is fed through the tunnel in steps.  At each step, the portion of the mRNA in the tunnel dictates which type of amino acid is allowed to enter the tunnel and bind to the nascent chain.  As the chain gets longer and eventually terminates, it folds into the protein coded for by the mRNA.  The ribosome molecule is itself composed of two main subunits - 50s and 30s - which come together to form the tunnel where the proteins are assembled.  We show the ribosome in both primal and complementary space views in Figure \ref{fig:ribosomeComp}.  As we begin to measure the complementary space model, we will be able to provide evidence for or against various hypotheses about the protein construction process such as whether there is enough room inside the tunnel for proteins to begin folding.

Complementary space also aids in answering the question of how the ribosome comes into its assembled state.  The larger subunit alone (PDB ID 1FFK) is made up of a long, coiled RNA strand, a short RNA strand, and dozens of proteins various types, as shown in Figures \ref{fig:ribosomeComp}a-c.  While all the proteins involved can and have been identified and labelled, the order in which they come together to form the subunit is unknown as video capture techniques do not exist for the requisite nanometer-resolution scale.

We have created a video of a plausible assembly sequence and analyze its accuracy via a complementary space method as follows.  The PDB entry 1PNY provides atom locations for the assembled ribosome molecule with tags identifying those atoms belonging to the various docked proteins.  We separate the atoms according to their tags and create meshed surface representations of each of the proteins and the coiled RNA strands; the resulting geometries are thus still fixed in space according to the PDB location information.  For each protein geometry, we search for vertices lying within four angstroms of the RNA geometry using the distance function $h_P$.  Triangles incident to these vertices are tagged as part of the ``contact region.''

We sum the areas of the triangles in the contact region for each of the 32 proteins.  We found that the contact regions with RNA vary greatly in size - from 324 $\text{\AA}^2$ to 7385 $\text{\AA}^2$.  The proteins with larger contact areas  bind to RNA first while the ones with smaller areas bind later as their access to RNA is partially blocked by other proteins.

\subsection{Virus models}
\begin{figure}
\centering
\includegraphics[width=\textwidth]{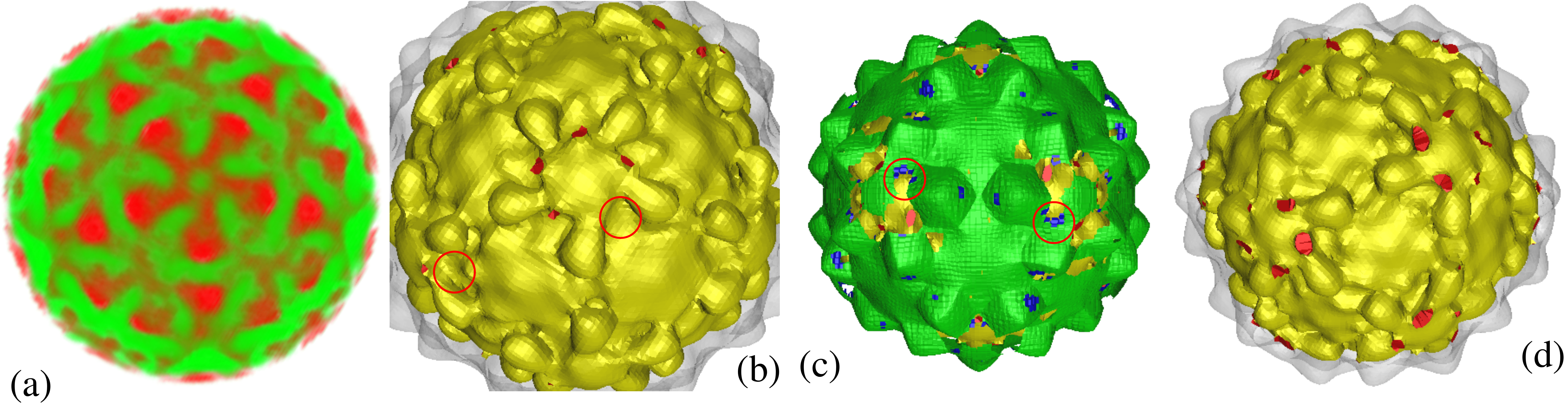}
\caption{Identification of ``thin'' regions in the primal space for the nodavirus dataset.
         \textbf{(a)} A volume rendering of the 3D image data.
         \textbf{(b)} Tunnels are detected for the initial selection of
             the isosurface. Note that in some places of 5-fold
             symmetry, only 4 mouths of the tunnel are present.
         \textbf{(c)} The thin regions (blue) are identified as subsets
             of the unstable manifolds of the index 1 saddles
             identified on the interior medial axis. The circles (red) in (b) and (c) indicate that places where
             the fifth mouth of the tunnel should be open indeed have thin regions.
         \textbf{(d)} The final selected isosurface has complementary
             space topology consistent with the inherent symmetry
             of the 3D density map.}
\label{fig:noda_thin}
\end{figure}

Viruses rely heavily on their geometry to infect cells and replicate their genome.  Since the goal of a virus is rapid reproduction, many viruses are made of identical subunits forming a highly symmetrical capsid shell, thereby minimizing the number of unique parts that must be synthesized.  Accordingly, an accurate model of a virus should have the same symmetries as the virus itself and complementary space can aid in detecting such symmetries.

The example we consider here is the nodavirus which infects certain types of freshwater fish.  A simple volume rendering is shown in Figure \ref{fig:noda_thin}a with the symmetry evident.  Selecting an isosurface from the range of possible values, however, presents a challenge as noise in the data often upsets the symmetry as is seen in Figure \ref{fig:noda_thin}b.  The problem in this case is not the presence of a small tunnel but the absence of one, both at the indicated area and elsewhere.  We therefore need a new type of complementary space visualization indicating ``thin regions'' where the surface comes close to a self intersection; these regions are candidates for a missing tunnel.

Remarkably, the distance function $h_P$ plays an important role here also.  The idea is to compute the interior medial axis of the surface and detect those portions lying very close to the surface, i.e. where $h_P$ is below some threshold $\gamma$.  For surfaces derived from image data, $\gamma$ should be set to the resolution of the imaging device as any smaller size features are probably unreliable and should be eliminated.  In practice, the medial axis is often noisy resulting in many erroneous thin regions, so we use instead two subsets of it ($U_1$ and $U_2$, described below) which are stable against small undulations on the surface.  The method works as follows.
\begin{enumerate}
\item We first approximate the interior medial axis using the Voronoi and Delaunay diagrams already computed for complementary space modeling.  The method for this is described in a previous paper \cite{GGB2007}.
\item Collect the point sets
    \begin{eqnarray*}
        C_{1,IM} &=& \{\text{index 1 saddles of $h_P$ on int. med. axis of $\Sigma$}\} \\
        C_{2,IM} &=& \{\text{index 2 saddles of $h_P$ on int. med. axis of $\Sigma$}\}
    \end{eqnarray*}
\item Compute the unstable manifolds of each of point in these two sets.  This results in a piecewise planar subset $U_1$ of the medial axis for points from $C_{1,IM}$ and a piecewise linear subset $U_2$ for points from $C_{2,IM}$.
\item At each Voronoi vertex lying on $U_1$ or $U_2$, compute the value of $h_P$.  This is given by the circumradius of the Delaunay tetrahedra dual to the Voronoi vertex.  If $h_P$ is smaller than $\gamma$, mark the region as ``thin'' and color differently for visualization.
\item Collect the interior maxima falling into the thin subsets of $U_1$ and $U_2$ and compute their stable manifolds.  The stable manifold creates a geometry for the missing tunnel.
\end{enumerate}

Figure \ref{fig:noda_thin} (c) shows the thin regions (blue patches) on the $U_1$ stable manifold (green) identified for the nodavirus model.  These can be selectively replaced by tunnels to capture the correct symmetry.  Alternatively, the presence of thin regions suggests a different isovalue may be more appropriate, such as the one shown in Figure \ref{fig:noda_thin} (d).

\subsection{Topological Consistency of Reduced Models}
\begin{figure}
\centering
\includegraphics[width=\textwidth]{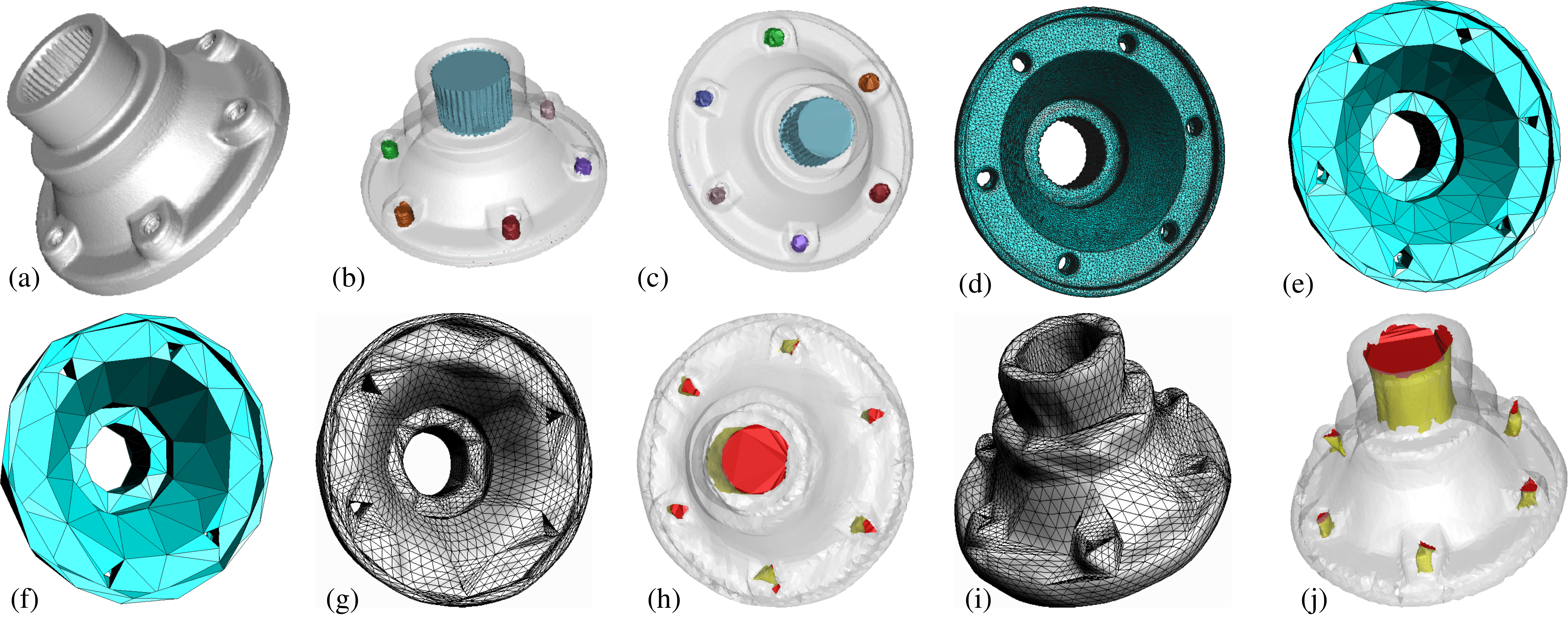}
\caption{Visualizations of the Carter dataset.  \textbf{(a)} A basic primal space visualization of the mechanical part.  \textbf{(b-c)} Complementary space features identified and visualized. \textbf{(d)} A visualization of the dense mesh representing the surface reveals that at 106,708 triangles, it is probably amenable to decimation.  \textbf{(e-f)} Using QSlim \cite{qslim}, the mesh is decimated to 1000 and then to 500 triangles. \textbf{(g-j)} After refinement and geometric improvement on the 500 triangle model, it appears from the primal space visualizations (g and i) that some of the tunnels have collapsed, a topological change.  Complementary space visualizations (h and j) reveal that in fact the tunnels are still present but with perturbed geometry.  Therefore, this reduced model has consistent topology with the original.}
\label{fig:carter}
\end{figure}

Model reduction or decimation is the process of removing geometrical information from a model while attempting to keep sufficient data for maintenance of important features.  This is used, for example, in coarse-grained models of proteins for electrostatic simulations \cite{BBH2007}.  Protein surfaces are often defined based on atomic positions and radii, obtained from the PDB.  For large proteins, a significant speed-up in computational time can be achieved by grouping atoms into clusters and treating the clusters as single atoms with an averaged radius.  Model reduction is also common for point-sampled surfaces such as CAD models and geometries acquired from three-dimensional scanners.  If points on the surface can be culled without dramatic effect on the shape of the surface, subsequent visualization and simulation pipelines will have reduced computational cost.

Complementary space visualization aids in determining when, if ever, the original topology is lost in progressive decimation.  Consider the industrial part model shown in Figure \ref{fig:carter}.  We use the software QSlim \cite{qslim} to decimate the model from 106,708 triangles to only 1000 and then only 500.  At 1000 triangles, the model has lost some geometrical precision, but still appears to have the same number of tunnels (\ref{fig:carter} e).  At 500 triangles, however, some of the tunnels appear to have collapsed (\ref{fig:carter} f).  We improve the geometry by refinement and smoothing but still cannot tell from primal space visualizations if the topology is correct (\ref{fig:carter} g and i).  Complementary space visualizations, however, quickly show that all tunnels are indeed present, albeit somewhat distorted (\ref{fig:carter} h and j).  From a topological standpoint, therefore, this reduction is consistent; the application context will determine if it is acceptable from a geometrical standpoint as well.

\subsection{Dynamic Deformation Visualization}
\label{sec:hemo}
\index{Time-dependent topology}
Complementary space aids in visualizing and quantifying dynamic deformations of models in addition to its aid for static models previously discussed.  The omnipresent consideration in a computer generated simulation of real movement is always whether the dynamics are realistically plausible.  In the context of molecular modeling, such considerations are especially difficult to formalize as current video technology cannot capture a molecule in vivo for comparison.  As a result, various techniques have been developed for automated animation of molecular models, including the popular method of Normal Mode Analysis (NMA) \cite{LSS1985,T2003}.

To determine whether the fluctuations simulated by these means have any functional significance to the molecule, we must be able to measure the extent of changes in particular features of the surface.  This is especially important in molecules which perform specific actions by modifying their complementary space features, such as the ribosome.  With a model of complementary space, we can measure the area of the mouth of a tunnel or pocket used in the various processes and compare the sizes before and after a conformational change.  This gives insight into the relative magnitude of different aspects of the shape reconfiguration; a seemingly significant deformation may only involve a small change in the size of a pocket mouth or vice versa.

To demonstrate the benefits of complementary space dynamics visualization, we consider the hemoglobin molecule.  Hemoglobin is the vehicle used to transport oxygen through the bloodstream.  A single hemoglobin molecule is made of four subunits, each of which can hold one oxygen molecule at its heme group.  This conformational change has been simulated by interpolation of PDB data for the bound and unbound states by our collaborators Drs. David Goodsell and Arthur Olson of the Molecular Graphics Laboratory at the Scripps Research Institute.  Using this sequence of time steps, we have generated videos of primal and complementary space dynamics.  We compare the visual differences in primal and complementary space further in Figure \ref{fig:hemVideo_viz}.  Interestingly, the complementary space has dynamically changing geometry, both in the region of the active site and deeper in the interior of the molecule.

\begin{figure}
\centering
$\begin{array}{cc}
\includegraphics[width=2.5in]{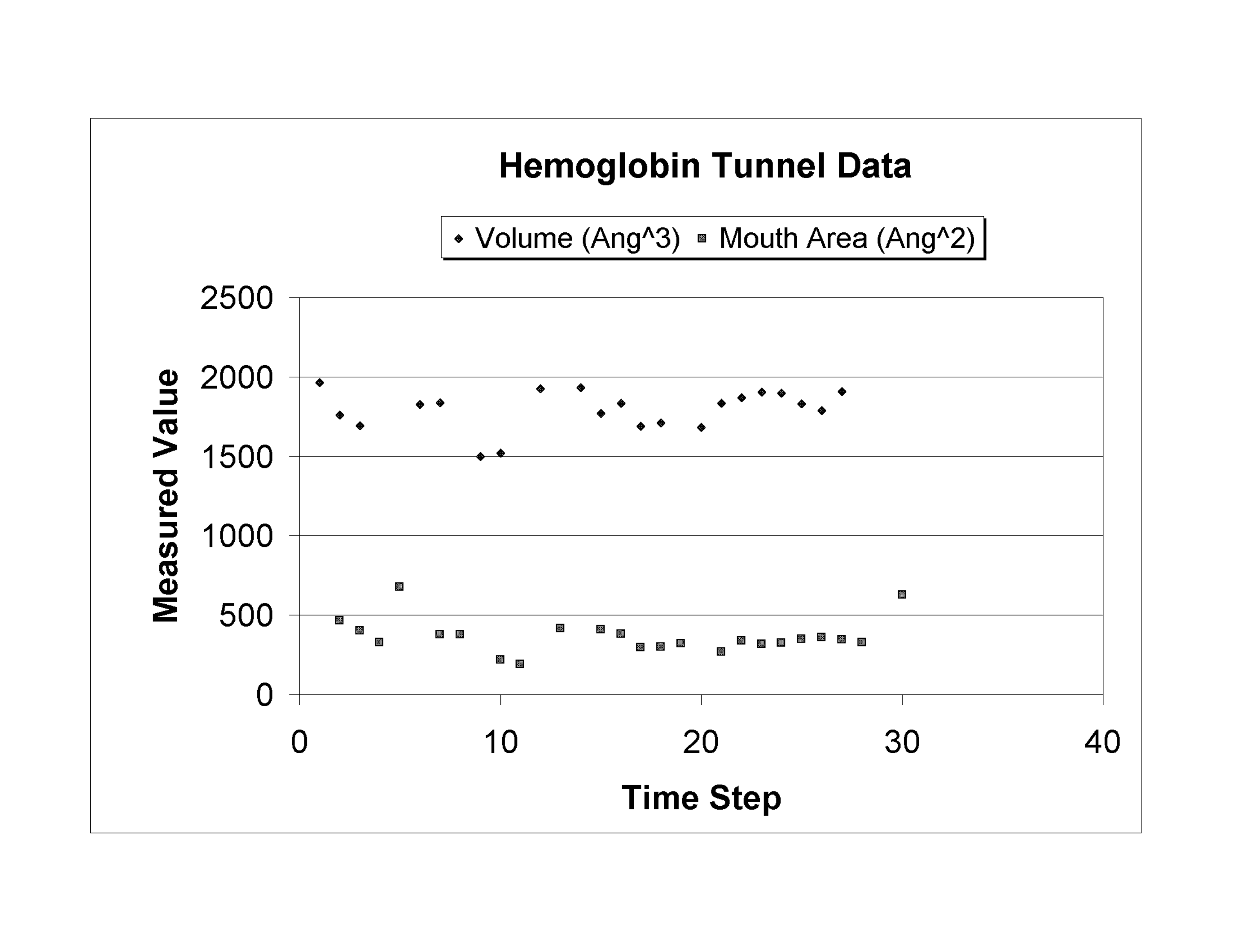} & \includegraphics[width=2.5in]{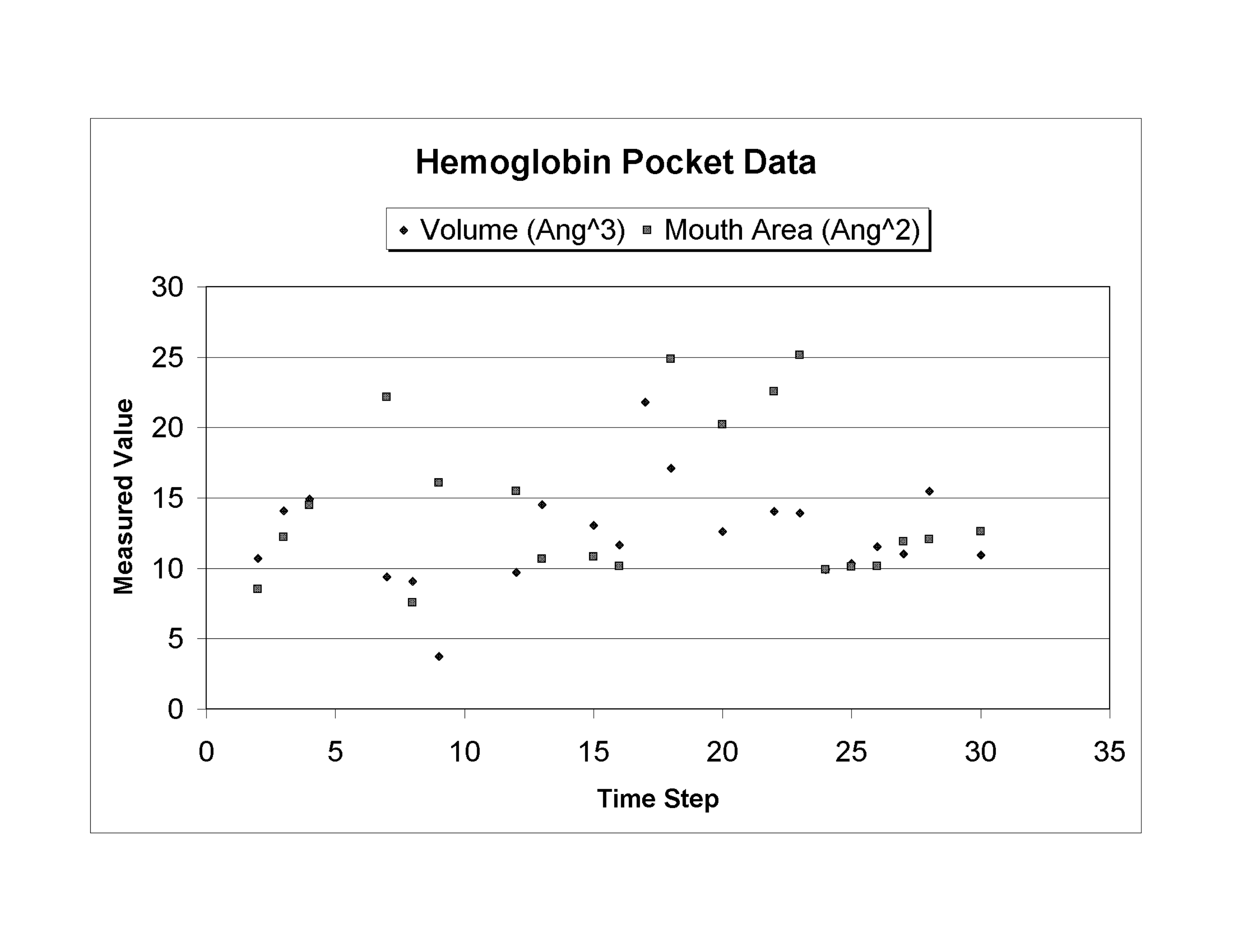}
\end{array}$
\caption{Quantification of the complementary space tunnel in the hemoglobin time step data.  The undulating nature of the two data series reflects the dynamic deformation hemoglobin undergoes while binding to oxygen.}
\label{fig:hem_tun_data}
\end{figure}

We also quantify this data by measuring the volume of the two main complementary space features of the molecule detected by \textsc{CompSpace}: a tunnel and a pocket.  For each feature, we compute the enclosed volume (in $\text{\AA}^3$) and total mouth surface area (in $\text{\AA}^2$).  We show the results in Figure \ref{fig:hem_tun_data}.  The tunnel data shows a dynamic change in enclosed volume over the time scale, with a max of 1963 and a min of 1498 $\text{\AA}^3$.  The total mouth area also varies widely, with a max of 679 and a min of 190 $\text{\AA}^2$.  The pocket data exhibits similar fluctuations.

\section{Conclusion}
\label{sec:conc}

Visualization of the complementary space of a geometrical model has the immediate impact of elucidating pockets, tunnels, and other subtle structural features.  We have shown in this paper that an explicit geometry of complementary space is often essential to allow for the measurement of certain aspects of the model such as tunnel mouth area or pocket volume.  These quantities can characterize the feasibility of models in biology or the precision of CAD-based models.  As we have discussed with the hemoglobin example, complementary space also plays a useful role in creating and analyzing dynamic visualizations.  Finally, as seen in the case of ribosome assembly, complementary space measurements can also guide the creation of primal space dynamics visualizations.

It is in this last vein of questions regarding assembly pathways that we intend to expand this work.  To identify likely assembly paths, we construct a graph whose nodes are the assembly parts in question; in the case of the ribosome, these are the RNA chain and the individual proteins which bind to it.  Edges exist between parts which are adjacent and edge weights are given by contact surface area.  Assembly order is based on relative affinity between parts and affinity is a function of contact area.  Hence, an assembly path relates closely to a maximally weighted spanning tree of the nodes.  We plan to elucidate this further in future work.

\textsc{Acknowledgments}  We would like to thank previous members of the CCV lab who worked on these projects, including Katherine Claridge, Vinay Siddavanahalli, and Bong-Soo Sohn.  This research was supported in part by NSF grants DMS-0636643, CNS-0540033 and NIH contracts R01-EB00487, R01-GM074258, R01-GM07308.

\bibliographystyle{abbrv}

\bibliography{tiv2009_book}
\end{document}